\documentclass[
 preprint,
 superscriptaddress,
 amsmath,amssymb,
 aps,
]{revtex4-2}

\usepackage{graphicx}
\usepackage{dcolumn}
\usepackage{bm}
\usepackage{hyperref}

\begin{document}

\title{Overlimiting Ion Transport and Reaction Limitations in Charged Porous Media}

\author{Arjun V. Yennemadi}
\thanks{These authors contributed equally to this work.}
\affiliation{Department of Chemical Engineering, Massachusetts Institute of Technology, Cambridge, Massachusetts 02139, USA}

\author{Junghyun Yoon}
\thanks{These authors contributed equally to this work.}
\affiliation{Department of Chemical Engineering, Massachusetts Institute of Technology, Cambridge, Massachusetts 02139, USA}

\author{Martin Z. Bazant}
\email{bazant@mit.edu}
\affiliation{Department of Chemical Engineering, Massachusetts Institute of Technology, Cambridge, Massachusetts 02139, USA}
\affiliation{Department of Mathematics, Massachusetts Institute of Technology, Cambridge, Massachusetts 02139, USA}

\date{\today}

\begin{abstract}
Electrochemical reaction rates are controlled by both interfacial charge-transfer kinetics and reactive-ion transport. In charged porous media, fixed charges enrich reactive counterions near pore walls, enabling transport beyond the classical diffusion limit via surface conduction (SC). Under overlimiting conditions, classical Butler--Volmer kinetics predict indefinitely increasing current with overpotential, contrasting with microscopic electron-transfer theories, which impose a finite reaction-limited current. Here, we couple the one-dimensional leaky membrane model to coupled ion--electron transfer (CIET) kinetics to examine the interplay between transport and reaction limitations. The limiting behavior is governed by the scaled surface charge \((\tilde{\rho}_s)\) and a Damköhler number \((Da)\) comparing reaction-limited and diffusion-limited currents. We derive analytical limiting-current expressions for neutral, positive, and negatively charged porous media, mapping underlimiting-to-overlimiting transitions in the \((Da,\tilde{\rho}_s)\) plane. By preventing reactive-ion depletion, SC restores polarization-curve sensitivity to charge-transfer kinetics that would otherwise be obscured by diffusion limitation. Fitting to published Cu electrodeposition data in charged AAO membranes yields \(Da\approx24\). CIET parameters fitted to AAO(\(-\)) also describe AAO(\(+\)) and predict a finite AAO(\(-\)) reaction limit beyond the applied voltage range. These results provide a framework for distinguishing transport-limited and reaction-limited responses in electrochemical systems and establish charged porous media as platforms to reveal electrochemical reaction-kinetic descriptors.

\end{abstract}

\maketitle

\section{Introduction}
Electrochemical processes are being increasingly utilized for energy storage \cite{goodenough_electrochemical_2014, yang_electrochemical_2011, ragupathy_electrochemical_2023, xiao_recent_2021}, selective separations \cite{alkhadra_electrochemical_2022, arges_current_2025, wu_recent_2022, kim_electrochemical_2021}, resource recovery \cite{kim_electrochemical_2021, battistel_electrochemical_2020, yao_electrochemical_2026,sood_electrochemical_2021, holmes_electrochemical_2025}, and environmental remediation \cite{chen_electrochemical_2004, muddemann_electrochemical_2019, miller_electrochemical_2023, sullivan_coupling_2021, renfrew_electrochemical_2020}. Across these applications, ions are often transported through porous media such as electrodes, membranes, or catalyst layers before undergoing charge-transfer reactions at electron-conducting interfaces. As a result, device performance is controlled not only by the intrinsic kinetics of the interfacial reaction, but also by the ability of the porous medium to deliver reactive ions to the interface. This coupling is especially important in charged porous materials, where fixed charges on pore walls can strongly reshape ion transport, as demonstrated in nanochannels \cite{zangle_theory_2010, nam_experimental_2015}, electrodeionization systems \cite{pourcelly_applications_2012, nikonenko_desalination_2014, alvarado_electrodeionization_2014, pan_development_2017, palakkal_advancing_2020}, and electrodialysis \cite{deng_overlimiting_2013, schlumpberger_scalable_2015, alkhadra_small-scale_2020, alkhadra_continuous_2022}. In reactive systems such as batteries and electrodeposition, these transport effects strongly modify Faradaic reaction rates; experiments have shown that the sign and magnitude of fixed charge can directly influence the current–voltage response, suppress or promote dendritic growth, and change deposit morphology \cite{han_over-limiting_2014, han_dendrite_2016, tikekar_stabilizing_2016, zhi_biomolecule-guided_2020, shao_regulating_2022, khoo_linear_2019}. A predictive description of such systems therefore requires a framework that connects ion transport through charged pores with a detailed description of electrochemical reaction kinetics.

The limiting cases of this coupled problem have been studied extensively \cite{newman_electrochemical_2021, bard_electrochemical_2022}. Classical models of ion transport to consuming interfaces describe diffusion, advection, and electromigration in neutral electrolytes using the Nernst–Planck equations, often under electroneutrality. These models relate the ionic current to the electric potential drop across the electrolyte (a proxy for the driving force), thereby providing the classical current–voltage response of a diffusion layer or neutral porous medium. For weak advection, as the applied driving force increases, these models predict a maximum \textit{diffusion-limited current}, which is reached when the reactive ion becomes depleted near the electrode in a phenomenon known as ion concentration polarization. In charged porous media, fixed charges on pore walls induce electric double layers, and the full pore-scale transport problem is described by the Poisson-Nernst-Planck theory \cite{mani_propagation_2009,zangle_theory_2010,mani_deionization_2011}. Reduced descriptions, such as the leaky membrane model \cite{dydek_nonlinear_2013}, avoid explicitly resolving these double layers by instead enforcing cross-sectional electroneutrality to retain the effect of the fixed charges on the cross-sectionally averaged ion concentrations. Under strong driving forces, these models have shown that fixed charges of opposite sign to the reacting ionic species can facilitate \textit{surface conduction} (SC), wherein ions transport along the electric double layers at the pore walls, bypassing the depleted bulk. Surface conduction has been well-studied in many early works \cite{von1905theorie, bikerman1935wissenschaftliche,urban_contribution_1935,deryagin1969theory, bikerman_electrokinetic_1940, overbeek1950quantitative}, but it was shown much later that SC can enable transport beyond the classical diffusion limit, allowing the medium to sustain \textit{overlimiting} currents (OLC) and deionization shocks \cite{mani_propagation_2009,zangle_theory_2010, mani_deionization_2011, dydek_overlimiting_2011}. Other physical mechanisms such as electroosmotic flow (EOF) \cite{dydek_overlimiting_2011, nam_experimental_2015, mani_deionization_2011} and the Rubinstein-Zaltzman electroosmotic instability (EOI) \cite{rubinstein_electro-osmotically_2000,zaltzman_electro-osmotic_2007, rubinstein_direct_2008}, and chemical mechanisms such as water splitting \cite{nikonenko_intensive_2010, andersen_current-induced_2012} and charge regulation \cite{andersen_current-induced_2012} can also facilitate OLC. The dominant physical mechanism for OLC is controlled by the pore aspect ratio and the surface charge, and SC generally dominates in long and thin pores (aspect ratios below $10^{-2}$) with moderate to high surface charge density \cite{dydek_overlimiting_2011}.

Separately, ion transport coupled to Faradaic reaction has been studied in the context of diffusion layers, neutral porous media, and porous electrodes \cite{parrish_current_1969,newman_porous-electrode_1975,biesheuvel_diffuse_2011,smith_na-ion_2016, yan_theory_2017}. These models typically combine Poisson–Nernst–Planck transport with Butler–Volmer (BV) kinetics at the reactive interface, capturing the transition between kinetic control at low driving force and transport limitation at high driving force. The full coupling between Faradaic reactions and fixed-charge-mediated ion transport was modeled by Khoo et al. \cite{khoo_theory_2018} to describe experiments on Cu electrodeposition and electrodissolution in charged nanoporous membranes by Han et al. \cite{han_over-limiting_2014, khoo_theory_2018}. For negatively charged membranes, the model predicts that with increasing overpotential, surface conduction continues to supply reactive Cu$^{2+}$ ions beyond the classical diffusion-limit, in agreement with experiment. Importantly, though, the model assumes Butler–Volmer (BV) kinetics for the two Faradaic reactions, and while BV kinetics can capture the current–voltage response over the voltage range studied in their work, the model predicts that increasing cell voltage can lead to indefinitely high currents.

This unphysical prediction arises because, in classical BV kinetics~\cite{butler_hydrogen_1936,erdey-gruz_zur_1930}, the reaction rate does not possess an intrinsic upper bound when the interfacial reactant concentration is fixed, for instance, to balance the fixed surface charge. In principle, though, beyond a certain overpotential, the transition from the oxidized to reduced state should become barrierless, eliminating the kinetic dependence on overpotential. This behavior is captured by microscopic electron-transfer theories, specifically Marcus theory for metal electrodes, which predicts curved Tafel behavior and a finite reaction-limited current at large overpotential~\cite{marcus_theory_1956,marcus_theory_1965,hush_electron_1999,chidsey_free_1991}. The polarization curve can therefore be limited either by ion transport through the porous medium or by the maximum interfacial charge transfer rate.

The recently developed coupled ion--electron transfer (CIET) theory describes Faradaic reactions on a two-dimensional landscape of excess free energy defined by an ion-transfer (IT) coordinate and an electron-transfer (ET) coordinate~\cite{bazant_unified_2023,fraggedakis_theory_2021,yoon_quantum_2026}. In the ion-coupled electron-transfer (ICET) limit, the IT free-energy barriers exceed the Marcus reorganization energy and the energy associated with the applied overpotential. Consequently, slow IT brings the reaction complex to the diabatic crossing, where ET occurs rapidly, and CIET reduces to BV kinetics, with the exchange current and symmetry factor linked to microscopic interfacial properties. By contrast, in the electron-coupled ion-transfer (ECIT) limit, the Marcus reorganization energy exceeds the IT barriers and the energy associated with the applied overpotential. Slow solvent or lattice reorganization then brings the diabatic electronic states into resonance, and the resulting Marcus--Hush--Chidsey (MHC)-type rate for metal electrodes retains an effective IT free-energy barrier in its prefactor. CIET therefore spans both limits and their transition while retaining the BV limit used in earlier overlimiting-current models~\cite{khoo_theoretical_2019,han_over-limiting_2014}. At sufficiently large formal overpotential, however, the kinetics cross into ECIT regime, where barrierless transitions produce a finite reaction-limited current.

Hence, in this work, we couple a one-dimensional leaky membrane model of ion transport to CIET reaction kinetics at the reactive boundary to understand the interplay between transport and reaction limitations. Arising from the analysis are two dimensionless groups of particular interest, namely, the Damk\"ohler number $Da$, representing the competition between the reaction-limited and diffusion-limited currents, and $\tilde{\rho}_s$, the scaled surface charge of the porous medium. We (1) derive limiting current expressions for neutral, positively charged, and negatively charged porous media, (2) map the transition between underlimiting and overlimiting current in terms of $Da$ and $\tilde{\rho}_s$, and (3) showcase how surface conduction in the overlimiting regime exposes the features of charge transfer kinetics that diffusion limitations would otherwise hide. We then apply the model to published Cu/AAO/Cu polarization curves of Han et al. \cite{han_over-limiting_2014} and examine the applied potential ranges over which BV and MHC expressions provide useful reduced descriptions of CIET kinetics.

\section{Theory}

\subsection{Mass Transfer Models}\label{sec2:mass_transfer}

We consider steady, one-dimensional transport of ions from a feed reservoir of salt concentration $c_0$ across a charged porous medium of length $L$, cross-sectional area $A$, porosity $\epsilon_p$, and tortuosity $\tau$, following an approach similar to the Leaky Membrane Model \cite{dydek_nonlinear_2013}. For simplicity, we consider a binary electrolyte (with ions $X^{z+}$ and $A^{z-}$, where $z$ is the ion valency), but we note that our approach can be easily extended to other electrolytes. The pore walls carry a surface charge density $q_s$, which we assume to be uniform along the cross section. The length scale characteristic of the pores, denoted by $h_p$, is defined as $\epsilon_p/a_p$, where $a_p$ is the pore surface area per volume of the porous medium. Then, electroneutrality at each cross section yields

\begin{equation}
     ze(c_X - c_A)\epsilon_p + q_sa_p = 0
\end{equation}

where $e$ is the elementary charge, and $c_X(x)$ and $c_A(x)$ refer to the concentrations of $X^{z+}$ and $A^{z-}$ in the pores at a given location $x$. The transport of the two ions across the medium is described by the 1D Nernst-Planck equations in the ideal (dilute) limit, as follows.

\begin{align}
     J_X = -D_X'\left(\frac{dc_X}{dx} + \frac{zc_X}{k_BT/e}\frac{d\phi}{dx}\right)
     \\
     J_A = -D_A'\left(\frac{dc_A}{dx} - \frac{zc_A}{k_BT/e}\frac{d\phi}{dx}\right)
\end{align}

Here, $D_i'=\frac{D_i\epsilon_p}{\tau}$ represents the effective diffusivity of each ion (different from the bulk diffusivity $D_i$), $\phi(x)$ is the electric potential, and $k_BT/e$ is the thermal voltage equal to 25.7 mV at standard conditions. $J_X$ and $J_A$ are the fluxes of each species, and species conservation demands that these must be constant across the axis at steady state.

At $x=L$, we consider a cathode at which $X^+$ is consumed via the reduction reaction $X^{z+} + ze^- \rightarrow X$, whose kinetics are described by the following general form.

\begin{equation}
     I \equiv zeAJ_X = zeAk c_X|_{L} \ f(\eta_f) \label{eq:4}
\end{equation}

Here, $I$ represents the observed current, $k \equiv k_{\rm{CIET}}$ is the reaction rate constant, and $c_X|_{L}$ is the concentration of $X^{z+}$ at the electrode. $\eta_f = V_e-\phi_L-E^{0,\prime}$ represents the formal overpotential, and for simplicity, we neglect the double layer at the cathode, which can be accounted for in a manner similar to previous studies \cite{chu2005electrochemical, yan_theory_2017, biesheuvel_diffuse_2011}. In this work, we consider the CIET reaction kinetic model, corresponding to a specified functional form for $f(\eta_f)$ that is described in detail in Section~\ref{sec2:CIET}. The cathode is assumed to block anions so that

\begin{equation}
     J_A = 0
\end{equation}

\begin{figure}[htbp]
    \centering
    \includegraphics[width=0.95\linewidth]{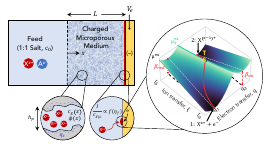}
    \caption{Schematic of the key physical processes captured by the continuum theory. Cations transport across a charged porous medium and are consumed via Faradaic reaction at the electrode at the far end. {The Faradaic reaction is described by coupled ion-electron transfer (CIET) kinetics, which captures reaction limitations by accounting for the energy barriers for oxidation and reduction in a single excess free energy landscape.}}
    \label{fig:fig1}
\end{figure}

A schematic of the modeled transport and reaction processes is shown in Fig.~\ref{fig:fig1}.

We remark here that without loss of generality, the analysis in this work also applies to the exact opposite case of anodes where anions are consumed and cations are blocked. Before proceeding, we scale the main quantities of interest as follows.
\[
\frac{x}{L}\rightarrow \tilde{x}
\]
\[
\frac{c}{c_0}\rightarrow \tilde{c}
\]
\[
\frac{\phi}{{k_BT}/{ze}}\rightarrow \tilde{\phi};
\qquad
\frac{\eta_f}{{k_BT}/{e}}\rightarrow \tilde{\eta}_f
\]
\[
\frac{I}{2zeAD'_Xc_0/L}\rightarrow \tilde{I}
\]

Doing so yields the following two dimensionless groups of particular interest in our work.

\[
\tilde{\rho}_s=\frac{q_s}{zec_0h_p}
\]

\[
{Da}=\frac{k}{2D_X'/L} = \frac{zeAkc_0}{2zeAD_X'c_0/L} \equiv \frac{I_{\rm{RL}}}{I_{\rm{DL}}}
\]

Here, $\tilde{\rho}_s$ is the ratio of the (volume-averaged) surface charge to the mobile ionic charge. $\tilde{\rho}_s$ measures how effectively the porous medium facilitates counterion conduction, which as Dydek et al. \cite{dydek_overlimiting_2011} note is different from the Dukhin number $Du$, which measures the relative importance of surface conductivity to bulk conductivity \cite{ delgado_measurement_2007, lyklema_surface_1998, bikerman_electrokinetic_1940}.  ${Da}$ is the Damköhler number which is the ratio between the Faradaic reaction-limited current ($I_{\rm{RL}}$) and the diffusion-limited current ($I_{\rm{DL}}$). Damköhler numbers commonly appear in chemical engineering problems coupling transport and reaction, such as catalysis and reactor design \cite{fogler_elements_1999}. Damköhler numbers closely related to our definition above also routinely appear in the analysis of reactive electrochemical systems such as batteries \cite{chakrabarti_modelling_2020, greco_limited_2021, majji_modeling_2023, maggiolo_solute_2020, wan_methodspotentialdependent_2021, pathak_scaling_2026}, electrosorption \cite{he_theory_2021, clarke_insights_2024}, electrodeposition \cite{khoo_linear_2019, fraggedakis_tuning_2020}, and electrocatalysis \cite{wan_methodspotentialdependent_2021, fu_heterogeneous_2015, halhouli_sensitivity_2016, lasia_hydrogen_1998}. 

Our set of nondimensionalized governing equations is thus given by

\begin{align}
    \tilde{I}
    &=
    -\frac{1}{2}\left(
    \frac{d\tilde{c}_X}{d\tilde{x}}
    +
    \tilde{c}_X\frac{d\tilde{\phi}}{d\tilde{x}}
    \right)
    =
    {Da}\,
    \tilde{c}_X|_{L}
    f(\tilde{\eta}_f)
    \label{eq:rxn_bc}
    \\
    0
    &=
    -\left(
    \frac{d\tilde{c}_A}{d\tilde{x}}
    -
    \tilde{c}_A\frac{d\tilde{\phi}}{d\tilde{x}}
    \right)
    \Leftrightarrow
    \tilde{c}_A=e^{\tilde{\phi}} \label{eq:anion_np}
    \\
    \tilde{c}_X
    &-\tilde{c}_A+\tilde{\rho}_s=0 \label{eq:en}
\end{align}

 Finally, at $\tilde{x}=0^-$, we consider a feed reservoir with electrolyte concentration $c_0$ and reference potential $\tilde{\phi}=0$. The Dirichlet boundary conditions at $\tilde{x}=0^+$ are obtained by applying the Donnan relations to the feed conditions as follows.

\begin{align}
    \tilde{c}_X|_{0^+}
    =
    \exp\left(-\tilde{\phi}_{0^+}\right)
    \\
    \tilde{c}_A|_{0^+}
    =
    \exp\left(\tilde{\phi}_{0^+}\right)
    \\
    \tilde{\phi}_{0^+}
    =
    \sinh^{-1}\left(\tilde{\rho}_s/2\right)
\end{align}

Lastly, in cases where an additional supporting electrolyte (say of type $Y^{z+} A^{z-}$) of concentration $c_{\rm{SE}}$ is added to the feed, such that $Y^{z+}$ is non-reactive, we modify Eqs. \ref{eq:anion_np} and \ref{eq:en} as follows
\begin{align}
    \tilde{c}_A
    &=(1 + \tilde{c}_{\rm{SE}})e^{\tilde{\phi}} 
    \\
    \tilde{c}_Y 
    &= \tilde{c}_{\rm{SE}}e^{-\tilde{\phi}}
    \\
    \tilde{c}_X
    &-\tilde{c}_A+ \tilde{c}_Y+\tilde{\rho}_s=0 
\end{align}
so that the additional ion concentrations are appropriately reflected in the electroneutrality relation.

With this modeling framework for ion transport, and having chosen a reaction kinetic model, we can determine the current ${I}$ for a given cathode voltage $V_e$. The surface charge density of the porous medium ($\tilde{\rho}_s$) is treated as a tunable design variable, while the Damköhler number $Da$ is treated as system-specific, since it incorporates information intrinsic to the cation's reaction-kinetic and transport properties. This framing presumes that the surface charge is a knowable property of the medium rather than another quantity to be fitted, and a mature body of interfacial measurements supports that presumption. The most direct route is electrokinetic methods. Specifically, streaming-potential and streaming-current measurements through the porous medium yield a zeta potential through the Helmholtz--Smoluchowski relation, from which, using an appropriate electric double layer model, one can estimate the surface charge density $q_s$ \cite{lyklema_surface_1998,delgado_measurement_2007,molina_streaming_1999, datta_characterizing_2010,nakamura_simultaneous_2012,saha_electrokinetic_2019,pendse_dynamic_2025}. 
Potentiometric acid--base titration returns the charge in absolute units, the net proton uptake of the amphoteric surface hydroxyls as a function of pH and ionic strength fixing both $q_s$ and the point of zero charge at which it vanishes \cite{noh_estimation_1989,lutzenkirchen_potentiometric_2012}. Interface-selective probes resolve the same charge in situ: second-harmonic generation reports the interfacial field through its field-induced contribution \cite{ong_polarization_1992,eisenthal_second_2006}, and synchrotron X-ray reflectivity resolves the counterion distribution that screens the wall to sub-angstrom precision \cite{fenter_mineralwater_2004}. The pH-resolved, absolute charge these methods return is the independent prior on $\tilde{\rho}_s$ that the identifiability analysis in Section \ref{sec:results_CIET} assumes.

\subsection{Reaction Kinetic Model: Coupled Ion-Electron Transfer} \label{sec2:CIET}

BV and MHC have each provided successful descriptions of charge transfer kinetics. BV reproduces the persistent Tafel behavior observed for many Faradaic reactions and has therefore become the standard kinetic law in continuum electrochemistry and electrochemical engineering~\cite{butler_hydrogen_1936,erdey-gruz_zur_1930,newman_electrochemical_2021,bard_electrochemical_2022}. By contrast, MHC derives the electrode reaction rate by integrating energy-dependent Marcus ET probabilities over the continuum of electronic states in the metal, weighted by their Fermi–Dirac occupations, while explicitly accounting for the collective solvent or lattice reorganization through the reorganization energy~\cite{marcus_theory_1956,marcus_theory_1965,chidsey_free_1991,henstridge_marcushushchidsey_2012}. This microscopic treatment predicts nonlinear Tafel behavior and a finite reaction-limited current at large overpotential.

CIET theory provides a unified kinetic framework for Faradaic reactions in which classical ion transfer and quantum mechanical electron transfer are treated within the same excess free-energy landscape \cite{bazant_unified_2023,fraggedakis_theory_2021,yoon_quantum_2026}. One coordinate represents IT, including ion displacement, changes in coordination, and desolvation; associated entropic constraints, such as site exclusion, enter through the transition-state activity. The other coordinate represents collective solvent or lattice reorganization that brings the oxidized and reduced diabatic states to equal free energy, thereby enabling isoenergetic ET. These states occupy distinct diabatic free-energy surfaces. ET becomes energetically allowed where the two surfaces intersect, and the minimum point along their intersection defines the CIET transition state. The IT free energies for the reduction and oxidation pathways are $\beta_{\rm red}$ and $\beta_{\rm ox}$, respectively, while $\lambda$ is the energy required to reorganize the environment from the equilibrium configuration of one electronic state to that of the other without transferring the electron. Changing $\eta_f$ shifts the two diabatic surfaces relative to one another and therefore changes both the location and height of the minimum CIET barrier. By retaining these coupled contributions, CIET can represent complex interfacial reactions without presupposing whether the measured kinetics are controlled by IT or ET.

The formal overpotential entering the CIET landscape is the quantity defined as
\begin{equation}
    \eta_f=V_e-\phi_L-E^{0,\prime}
\end{equation}
where $V_e$ is the electrode potential, $\phi_L$ is the electrolyte potential at the reaction plane, and $E^{0,\prime}$ is the formal potential of the reaction. Because CIET kinetics describes a single CIET event, all expressions in this section are written for one-electron transfer, and the corresponding formal overpotential is scaled by the thermal voltage $k_BT/e$, while the IT barriers and reorganization energy are scaled by the thermal energy $k_BT$, as follows.
$$\tilde{\beta}_{\rm red/ox}=\frac{\beta_{\rm red/ox}}{k_BT},\qquad \tilde{\lambda}=\frac{\lambda}{k_BT}.$$

For the linear IT landscapes used to get analytical CIET expression, the asymmetry is summarized by
\begin{equation}
\alpha=\frac{\beta_{\rm red}}{\beta_{\rm red}+\beta_{\rm ox}},
\end{equation}
where $\alpha$ is the cathodic symmetry factor.

In the IT-limited regime, the IT barriers are large compared with the reorganization energy and the applied driving force,
\begin{equation}
\beta_{\rm red},\beta_{\rm ox}\gg \lambda,e|\eta_f|,k_BT.
\end{equation}
In this limit, CIET kinetics reduces to the ICET expression. For continuum-scale reaction fitting, the reduction expression can be written in the same functional form as a BV equation,
\begin{equation}
I_{\rm ICET,red}=eAk_{\rm ICET} c_X|_{L} \exp(-\alpha\tilde{\eta}_f).
\end{equation}
Here, $A$ is the reactive electrode area, $k_{\rm ICET}$ is the ICET kinetic prefactor, and $\alpha$ is the cathodic charge transfer coefficient.

This expression is algebraically equivalent to the BV equation used in the fitting procedure. This equivalence, however, should be understood as a model-reduction statement rather than a statement that ICET is simply empirical BV kinetics. In the classical BV model, the prefactor is usually treated as an empirical exchange-current parameter. In CIET, by contrast, $k_{\rm ICET}$ contains microscopic information from the coupled ion-electron transfer event, including the quantum mechanical ET prefactor, the IT transition-state activity, the interfacial coverages or activities, and the geometry of the IT free-energy landscape~\cite{bazant_unified_2023}. Therefore, if the relevant microscopic quantities, such as electronic coupling, density of states, reorganization energy, transition-state activity, and interfacial free-energy landscape, were available from quantum simulations or molecular calculations, ICET could in principle be used for rate prediction rather than only empirical fitting.

The ICET expression is physically appropriate only when the voltage window remains inside the IT-limited Tafel regime, i.e., before the applied driving force becomes comparable to the IT barriers. This point is important because the physical IT picture underlying BV kinetics implies that the activation barrier should eventually be strongly reduced or removed at sufficiently large formal overpotential. The BV equation itself does not contain this barrierless transition. Instead, it simply continues the exponential Tafel branch without an intrinsic upper bound. Therefore, applying the BV equation outside its asymptotic range can unphysically predict indefinitely increasing reaction rates, whereas CIET predicts a transition away from the ICET regime once the applied driving force becomes comparable to the IT energy barriers.

In the ET-limited regime, the reorganization energy is large compared to the IT barriers and the applied driving force,
\begin{equation}
\lambda\gg \beta_{\rm red},\beta_{\rm ox},e|\eta_f|,k_BT.
\end{equation}
In this limit, CIET kinetics reduces to the electron-coupled ion transfer (ECIT) expression. For a metal electrode, integrating over the Fermi distribution of electron energies results in a Marcus-Hush-Chidsey (MHC)-type reduction rate~\cite{zeng_simple_2014,bazant_unified_2023,fraggedakis_theory_2021},
\begin{equation}
I_{\rm ECIT,red}=\frac{eAk_{\rm ECIT} c_X|_{L}}{2(1+\exp(\tilde{\eta}_f))}\operatorname{erfc}\left(\frac{\tilde{\lambda}-\sqrt{1+\sqrt{\tilde{\lambda}}+\tilde{\eta}_f^2}}{2\sqrt{\tilde{\lambda}}}\right).
\end{equation}
For continuum-level fitting, this expression shares the functional form of the reduced MHC model through an operational rather than conceptual equivalence. ECIT captures the reorganization-energy-dominant limit of CIET where IT participates in the elementary step while ET limits the rate, departing from the classical MHC picture of outer-sphere ET at a fixed ionic position~\cite{bazant_unified_2023}. The prefactor $k_{\rm ECIT}$ accounts for IT contributions, interfacial activities or coverages, electronic coupling, density of states, and reorganization effects. 

Thus, the MHC-type expression is appropriate when electron transfer and solvent or nuclear reorganization dominate the kinetic barrier. It becomes difficult to justify when experimentally observed currents maintain a nearly linear Tafel dependence over a wide overpotential range because Marcus-type ET kinetics generally produces curvature in Tafel plots outside the small-overpotential regime. In such cases, the ICET regime may provide the more appropriate reduced description, while the full CIET expression is needed when neither IT nor ET limitation is cleanly dominant.

While the two limiting expressions provide physical insight, the continuum model in this work uses a uniformly valid CIET expressions~\cite{bazant_unified_2023}. For the reduction branch, the reaction current is written as
\begin{equation}
I_{\rm CIET,red}=eAk_{\rm CIET}c_X|_{L} M(\tilde{\eta}_f)\mathcal{R}_{\rm red}(\tilde{\eta}_f;\tilde{\beta}_{\rm ox},\tilde{\beta}_{\rm red},\tilde{\lambda},\alpha) \equiv eAkc_{X}|_{L} \ f(\eta_f).
\end{equation}
Here, $M(\tilde{\eta}_f)$ is the asymptotic matching factor that smooths the transition between the IT-limited ICET branch and the ET-limited ECIT branch. In the present implementation, the reduction expression is evaluated using the ICET form before the IT barrier is exhausted and the ECIT form after the applied driving force reaches the oxidation-side IT barrier:
\begin{equation}\label{eq:ciet_kinetics}
\mathcal{R}_{\rm red}=
\left\{
\begin{aligned}
&-\tilde{\eta}_f<\tilde{\beta}_{\rm ox}:
\\[-0.2em]&\quad \displaystyle\frac{\exp[-\alpha(\tilde{\beta}_{\rm ox}+(1-\alpha)\tilde{\lambda})]}{\sqrt{4\pi\alpha(1-\alpha)\tilde{\lambda}}}\exp(-\alpha\tilde{\eta}_f),
\\[0.8em]
&-\tilde{\eta}_f\geq\tilde{\beta}_{\rm ox}:
\\[-0.2em]&\quad \displaystyle\frac{\exp[-\alpha(1-\alpha)\tilde{\lambda}]}{\sqrt{4\pi\alpha(1-\alpha)\tilde{\lambda}}}+\frac{1}{2}\Biggl[\operatorname{erf}\!\left(\sqrt{\alpha(1-\alpha)\tilde{\lambda}}\right)\\[-0.2em]
&\quad \displaystyle-\operatorname{erf}\!\left(\frac{2(1-\alpha)\tilde{\lambda}+\tilde{\beta}_{\rm ox}+\tilde{\eta}_f}{2\sqrt{\tilde{\lambda}(1-\alpha)/\alpha}}\right)\Biggr].
\end{aligned}
\right.
\end{equation}
The matching factor is
\begin{equation}
M(\tilde{\eta}_f)=M_{\infty}+\frac{M_0(\tilde{\eta}_f)-M_{\infty}}{\left[1+\exp\left(-\frac{\tilde{\eta}_f+\tilde{\beta}_{\rm ox}}{\tilde{\lambda}}\right)\right]\left[1+\exp\left(\frac{\tilde{\eta}_f-\tilde{\beta}_{\rm red}}{\tilde{\lambda}}\right)\right]},
\end{equation}
with
\begin{equation}
M_0(\tilde{\eta}_f)=\frac{\exp[-(1-\alpha)(\alpha\tilde{\lambda}+\tilde{\beta}_{\rm red})]\exp[(1-\alpha)\tilde{\eta}_f]}{\sqrt{4\pi\alpha(1-\alpha)\tilde{\lambda}}},
\end{equation}
and
\begin{equation}
M_{\infty}=\left[\frac{1}{2}\left(1+\operatorname{erf}\left(\sqrt{\alpha(1-\alpha)\tilde{\lambda}}\right)\right)+\frac{\exp[-\alpha(1-\alpha)\tilde{\lambda}]}{\sqrt{4\pi\alpha(1-\alpha)\tilde{\lambda}}}\right]^{-1}.
\end{equation}
This matched expression is the kinetic source model used to generate the synthetic OLC data. To relate this to the observed current of Eq.~\ref{eq:4}, we assume throughout that one CIET process is rate-determining and that all remaining steps are fast enough to remain quasi-equilibrated, so that no intermediate accumulates. Under this assumption, 
\begin{equation}
I=zI_{\rm CIET,red}=zeAkc_{X}|_{L} \ f(\eta_f),
\label{eq:I_from_CIET}
\end{equation}
where $z$ here enters only as the stoichiometric factor converting the rate of the rate-determining elementary step into the measured current. If the elementary steps instead have comparable rates, no single $f(\eta_f)$ of this form exists and an explicit multi-step reaction network would be required, which is outside the present scope.

\section{Results}
\subsection{Overlimiting Current Regime}\label{sec:results_OLC}
In this section, we will analyze the conditions under which the system yields a limiting current exceeding diffusion limitations. To begin, we define the limiting current as
\begin{equation}
    I_{\rm{lim}} \equiv \lim_{V_e\to-\infty} I(V_e)
\end{equation}

representing the current at extremely high cathodic voltage. The diffusion-limited current ($I_{\rm{DL}}$) is defined as the limiting current obtained when ion transport across a neutral porous medium ($\rho_s = 0$) is much slower than the Faradaic reaction at the cathode (i.e., $Da\rightarrow \infty$). In this limit, solving the model yields
\begin{equation}
    I_{\rm{lim}} =  \frac{2zeAD_X'c_0}{L} \equiv I_{\rm{DL}}  
\end{equation}
which as we mentioned previously, also serves as our scale for current. This current is obtained when the solution near the electrode is depleted of ions as the demand from reaction becomes very high, which is also known as concentration polarization. 

\begin{figure}[htbp]
    \centering
    \includegraphics[width=0.98\linewidth]{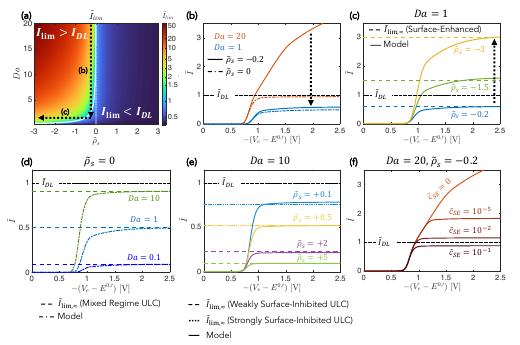}
    \caption{(a) Heat map showing the limiting current (scaled by the diffusion-limited current $I_{\rm{DL}}$) for combinations of $Da$ and $\tilde\rho_s$, considering CIET kinetics with parameters $\tilde{E}^{0,\prime} = -10, \ \tilde{\beta}_{\rm ox} = 30, \ \tilde{\lambda} = 15, \ \alpha  =0.8$. A nonlinear colorbar is applied to highlight the parameter space where the system transitions from underlimiting ($I_{\rm{lim}}<I_{\rm{DL}}$) to overlimiting ($I_{\rm{lim}}>I_{\rm{DL}}$), with the solid white curve indicating $I_{\rm{lim}}=I_{\rm{DL}}$. (b) {Current-voltage curves for reactions with high $Da$ (red-orange) and low $Da$ (blue) in negatively charged (solid lines) and neutral (dash-dot) porous media}. The change in the solid lines corresponds to movement along the arrow marked (b) in the heat map. (c) Current-voltage curves for reactions with the same $Da =1$ in increasingly negatively charged porous media, moving along the arrow marked (c) in the heat map. {The dashed lines are predictions of $\tilde{I}_{\rm{lim}}$ using the analytical approximations in Eq. \ref{eq:Ilim_neg_rhos}.} (d) Current-voltage curves for reactions with differing $Da$ in neutral porous media, along with $\tilde{I}_{\rm{lim}}$ predicted using Eq. \ref{eq:diff_lim}. (e) Current-voltage curves for reactions with $Da =10$ in increasingly positively charged porous media (solid lines), along with $\tilde{I}_{\rm{lim}}$ predicted using Eq. \ref{eq:Ilim_pos_rhos_small} (thick dotted lines) and Eq. \ref{eq:Ilim_pos_rhos_large} (dashed lines). (f) Effect of adding increasing amounts of supporting electrolyte for a reaction with $Da = 20$ in a negatively charged porous medium.}
    \label{fig:2}
\end{figure}

Next, we define \textit{overlimiting} current (OLC) as the case of $I_{\rm{lim}} > I_{\rm{DL}} \leftrightarrow \tilde{I}_{\rm{lim}} > 1$ and \textit{underlimiting} current (ULC) as $I_{\rm{lim}} < I_{\rm{DL}} \leftrightarrow \tilde{I}_{\rm{lim}} < 1$. Fig. \ref{fig:2}(a) shows a heat map of the (scaled) limiting current obtained by solving the model for various combinations of the two key system parameters $Da$ and $\tilde{\rho}_s$, considering monovalent ions ($z = 1$) and CIET reaction kinetics with representative parameters $\tilde{E}^{0,\prime} = -10, \ \tilde{\beta}_{\rm ox} = 30, \ \tilde{\lambda} = 15, \ \alpha  =0.8$. Our first observation is that neutral and positively charged porous media always yield underlimiting currents in alignment with the classical result \cite{dydek_nonlinear_2013}, and their corresponding I-V curves are shown in Fig. \ref{fig:2}(d) and (e). Negatively charged porous media are seen to enable overlimiting currents, however, that is only possible for the combinations of $Da$ and $\tilde{\rho}_s$ that lie above the white curve in Fig. \ref{fig:2}(a). As an example, the current-voltage curve for the case of high $Da= 20$ (effectively mapping to Butler-Volmer kinetics in this voltage range) is illustrated by the red-orange curve in Fig. \ref{fig:2}(b). In such systems, the negative surface charge allows cations to transport along the electric double layers formed at the pore walls (the SC mechanism), serving as a shunt resistance that bypasses the deionized bulk, allowing cations to exceed diffusion limitations. Theoretically, with no reaction limitations, such as for Butler-Volmer kinetics, one can obtain indefinitely high currents with increasing cathode voltage. However, as we show here, when reaction limitations are considered with CIET kinetics, the system always achieves a finite limiting current, which, depending on $Da$ and $\tilde{\rho}_s$, may be over- or underlimiting. Indeed, if reaction limitations are very strong, negatively charged porous media yield underlimiting current, as shown by the blue curve in Fig. \ref{fig:2}(b). Interestingly, though, our model predicts that even for systems with intrinsically slow kinetics (small $Da$), it is possible to bring the system into the overlimiting regime by increasing the negative surface charge, as seen in Fig. \ref{fig:2}(c). Thus, $\tilde{\rho}_s$ and $Da$ control the strength of surface conduction and the extent of reaction limitations, respectively, and together, these parameters determine whether the system is under- or overlimiting. 

As a qualification, however, we highlight here the effect of adding a supporting, non-reactive electrolyte to the system, considering the simple case of a monovalent binary electrolyte $Y^+ A^-$. For underlimiting cases ($\tilde{\rho}_s \geq0$), adding supporting electrolyte affects the I-V characteristics of the system only weakly. However, as shown in Fig. \ref{fig:2}(f), systems that would otherwise yield overlimiting current as per the heat map in Fig. \ref{fig:2}(a) can become underlimiting if excess supporting electrolyte is added. This is because the inert cations (here, $Y^+$) continue to remain in electrochemical equilibrium as the cathode voltage (and hence, local electrolyte potential) become more negative. Eventually, the inert cations replace all the reactive cations from the electric double layers close to the electrode, leading to $X^+$-ion concentration polarization and a plateauing of the I-V curve. This effect of supporting salts was experimentally observed in a recent study by Hong et al. where our collaborators found that a polyanion-coated polycarbonate membrane did not facilitate shock electrodeposition of cobalt, despite the negative surface charge --- here, our theory showed that diffusion limitations could be attributed to the presence of non-reactive Na$^+$ ions in their system~\cite{hong_selective_2026}.

\subsection{Analytical Approximations for the Limiting Current} \label{sec:approx}
For neutral media ($\tilde{\rho}_s = 0$), we can solve the model to yield the exact analytical result below.
\begin{equation} \label{eq:diff_lim}
    \tilde{I}_{\rm{lim}} = \frac{Da}{ 1 + Da} \ \leftrightarrow \ {I}_{\rm{lim}}^{-1} = \left(\frac{2zeAD_X'c_0}{L}\right)^{-1} + (zeAkc_0)^{-1} = I_{\rm{DL}}^{-1} + I_{\rm{RL}}^{-1}
\end{equation}

Fig. \ref{fig:2}(d) overlays the predictions from the above relation on the I-V curves for systems with differing $Da$. The above relation reflects the intuition that the total (limiting) resistance is a series combination of the transport resistance and the reaction resistance. The slower process among ion transport and reaction dominates the overall limiting current.

For positively charged porous media, we can approximate the limiting current by assuming  that near the electrode, cations are fully depleted and anions entirely balance the surface charge ($\tilde{c}_A|_L \approx \tilde{\rho}_s$), so that 
\begin{align}
\tilde{I}_{\rm lim}
&\approx
\frac{Da}{1+Da}\left[\exp\left(-\sinh^{-1}\left(\frac{\tilde{\rho}_s}{2}\right)    \right)+\frac{\tilde{\rho}_s}{2}\left(\ln \tilde{\rho}_s-\sinh^{-1}\left(\frac{\tilde{\rho}_s}{2}\right)\right)
\right], \label{eq:Ilim_pos_rhos}
\\
\intertext{where, for $\tilde{\rho}_s\to 0^+$,}
\tilde{I}_{\rm lim}
&\approx
\left(\frac{Da}{1+Da}\right)\left[1+\frac{1}{2}\left(\tilde{\rho}_s\ln\tilde{\rho}_s-\tilde{\rho}_s\right)+O\!\left(\tilde{\rho}_s^2\right)
\right], \label{eq:Ilim_pos_rhos_small}
\\
\intertext{and for $\tilde{\rho}_s\to\infty$,}
\tilde{I}_{\rm lim}
&\approx
\left(\frac{Da}{1+Da}\right)\left[\frac{1}{2\tilde{\rho}_s}+O\!\left(\frac{1}{\tilde{\rho}_s^3}\right)\right], \label{eq:Ilim_pos_rhos_large}
\end{align}

Fig. \ref{fig:2}(e) compares predictions from the approximations above to the plateau in the model-predicted I-V curves. We see that for positively charged media, the limiting current is always smaller than that for neutral porous media, and it further reduces by increasing the positive charge on the porous medium.

For negatively charged porous media, we obtain the result below.
\begin{subequations}\label{eq:Ilim_neg_rhos}
\begin{align}
\tilde{I}_{\rm lim}
&=
\left(\frac{Da}{1+Da}\right)\left[1-\frac{\tilde{\rho}_s(1+Da)}{2}\right]+O\!\left(\tilde{\rho}_s^2\right),
\qquad
0\leq-\tilde{\rho}_s \leq \frac{2}{1+Da},
\label{eq:Ilim_neg_rhos_small}
\\
\textrm{and } \tilde{I}_{\rm lim}
&=
Da\,(-\tilde{\rho}_s)+O\!\left(\frac{1}{-\tilde{\rho}_s}\right),
\qquad \qquad \qquad \qquad 
-\tilde{\rho}_s \geq \frac{2}{1+Da}.
\label{eq:Ilim_neg_rhos_large}
\end{align}
\end{subequations}
where these two limiting cases may be unified using a uniformly valid Padé approximation, as follows.
\begin{equation} 
    \tilde{I}_{\rm{lim}} = Da(-\tilde{\rho}_s) + \frac{Da}{1+Da}\left(\frac{1}{1+(-\tilde{\rho}_s)(1+Da)/2} \right) 
    \label{eq:Ilim_neg_rhos_uniform}
\end{equation} 

\begin{figure}[htbp]
    \centering
    \includegraphics[width=\linewidth]{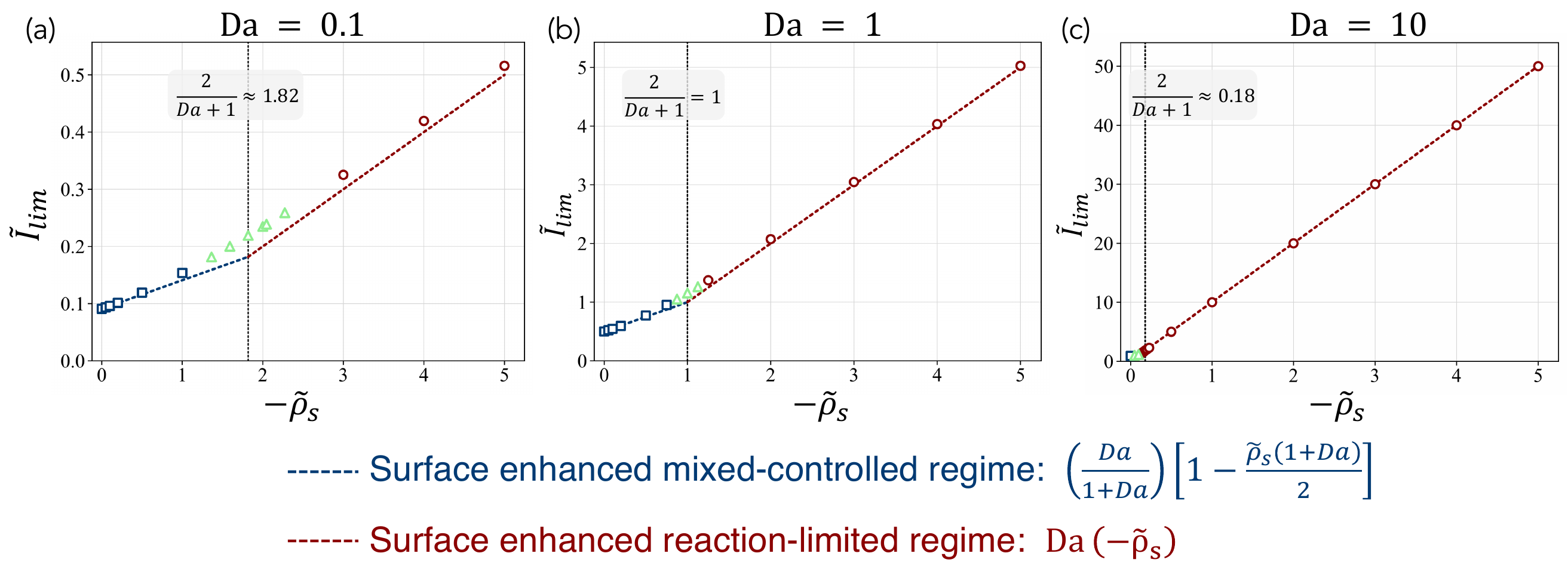}
    \caption{Analytical limiting-current estimates versus simulation across the ULC-to-OLC transition for (a) $Da=0.1$, (b) $Da=1$, and (c) $Da=10$. Markers: limiting currents of the full CIET reaction--transport model, with additional operating points placed near $2/(Da+1)$ to resolve the transition. Blue dashed line: weak-charge ULC law, Eq.~\eqref{eq:Ilim_neg_rhos_small}, drawn up to the predicted crossover; red dashed line: OLC law, Eq.~\eqref{eq:Ilim_neg_rhos_large}, drawn beyond it. Blue squares mark points reproduced within $10\%$ error by the ULC branch and red circles mark points reproduced within $10\%$ error by the OLC branch ; green triangles mark points where both approximations deviate from the simulation by more than $10\%$. The vertical dashed line is the predicted handoff $-\tilde{\rho}_s=2/(Da+1)$ of Eq.~\eqref{eq:Ilim_neg_rhos}, where the simulation exceeds the common analytic value $\tilde{I}_{\rm lim}=Da/(1+Da)$ by $17\%$, $13\%$, and $0.02\%$ for $Da=0.1$, $1$, and $10$, respectively.}
    \label{fig:analytical_estimate}
\end{figure}

Predictions of the limiting current using the approximations in Eq. \ref{eq:Ilim_neg_rhos} are shown as dashed lines in Fig. \ref{fig:2}~(c) for one case of $Da = 1$. Fig. \ref{fig:analytical_estimate} additionally compares the analytical approximations above to the model predicted $\tilde{I}_{\rm{lim}}$ for a range of values of $\tilde{\rho}_s$ and $Da = 0.1, \ 1, \textrm{ and }10$, illustrating that the simple approximations work well for most surface charge values, only underpredicting $\tilde{I}_{\rm{lim}}$ (at most by 17\%) close to the crossover point $-\tilde{\rho}_s \sim 2/(1+Da)$. These relations reinforce the idea that over- or underlimiting currents arise from the coupling of properties both intrinsic and extrinsic to the system. For instance, with especially slow kinetics (small $Da$), the system could remain below the diffusion limit even with a negatively charged porous medium if the magnitude of surface charge is small. However, Eq. \ref{eq:Ilim_neg_rhos_large} shows that if the porous medium is designed to have large enough negative surface charge, the boost in surface conduction can theoretically bring the system into the overlimiting regime. Surface conduction is one mechanism to enable OLC, but as described earlier, other mechanisms can also sustain overlimiting currents. Physical mechanisms such as electroosmotic flow and electroosmotic instability become important in larger pores or near bare electrodes, while chemical mechanisms such as charge regulation and water splitting can arise in systems with strong pH gradients coupled to Faradaic reactions. In future work, it will be important to determine how these mechanisms interact with reaction limitations, and whether enhancing them can similarly allow overlimiting currents in systems with intrinsically slow kinetics.

Furthermore, as illustrated in Fig. \ref{fig:2}(d), the shape of the I-V curve is qualitatively similar in the two mechanistically distinct cases of (i) fast reaction, slow transport (i.e., large $Da$, such as 10) and (ii) slow reaction, fast transport (i.e., small $Da$, such as 0.1), when the porous medium is neutral (equivalently, in the case of a diffusion layer over bare active surface). In other words, if the transport properties of the ion are not exactly known, which could be due to uncertainties in, say, the effective diffusivity or the diffusion film thickness, it is difficult to draw conclusions on the dominant resistance (transport or reaction) contributing to the plateau in the I-V curve. To this end, our theory would suggest placing a porous medium of known negative surface charge before the active area of reaction and measuring the limiting current. Specifically, we can use Eqs. \ref{eq:diff_lim} and \ref{eq:Ilim_neg_rhos} to write 
\begin{equation}
\frac{{I}_{\rm lim}(\tilde{\rho}_s<0)}
     {{I}_{\rm lim}(\tilde{\rho}_s=0)}
=
\begin{cases}
\displaystyle
1-\frac{\tilde{\rho}_s(1+Da)}{2}+O\!\left(\tilde{\rho}_s^2\right),
&
0\leq-\tilde{\rho}_s \leq \dfrac{2}{1+Da},
\\[10pt]
\displaystyle
(1+Da)(-\tilde{\rho}_s)+O\!\left(\dfrac{1}{-\tilde{\rho}_s}\right),
&
-\tilde{\rho}_s \geq \dfrac{2}{1+Da}.
\end{cases}
\label{eq:Ilim_ratio_negative_rhos}
\end{equation}
Thus, data on the change in the limiting current can be used with the relation above to estimate the Damköhler number $Da$. {As discussed in the following section, this procedure allows improved identifiability of $Da$ as opposed to fitting to the I-V profile at low voltage}. Since $Da$ represents the ratio of the reaction-limited current to diffusion-limited current, the value of $Da$ allows identification of the dominant resistance in the system: if the estimated $Da\gg1$, the transport resistance dominates, whereas if $Da\ll 1$, reaction resistance dominates.

\subsection{Charge transfer kinetic analysis with overlimiting current} \label{sec:results_CIET}

Under diffusion limitation, the measured current becomes weakly sensitive to the charge transfer parameters.  For $\tilde{\rho}_s=0$, the polarization curve approaches the transport-controlled plateau $\tilde{I}_{\rm lim}=Da/(1+Da)$ (Eq.~\ref{eq:diff_lim}), which tends to unity for $Da\gg1$. The interfacial concentration can then adjust with the kinetic parameters while the observed current remains near the same plateau, causing the partial sensitivities $\partial\tilde{I}/\partial\theta$, with $\theta\in\{k,\tilde\beta_{\rm ox},\tilde\lambda,\alpha\}$, to become small. Negative surface charge sustains cation transport after bulk depletion. Deep in the overlimiting regime ($-\tilde{\rho}_s\gg2/(1+Da)$), the observed current approaches its saturation value, giving $\tilde{I}_{\rm lim}=Da(-\tilde{\rho}_s)$ (Eq.~\ref{eq:Ilim_neg_rhos_large}). Thus, when $\tilde{\rho}_s$ is characterized independently, the plateau current is proportional to the interfacial charge transfer rate constant. The portion of the polarization curve at overlimiting current also retains the curvature associated with the charge transfer law and thereby restores sensitivity of the cell current to the interfacial kinetics.

We evaluated this effect at $Da=50$ by comparing the nearly diffusion-limited uncharged case, $\tilde{\rho}_s=0$, with two negatively charged cases capable of overlimiting current, $\tilde{\rho}_s=-0.05$ and $-0.2$. All calculations used the same prescribed CIET parameters ($\tilde{\beta}_{\rm ox}=8$, $\tilde{\lambda}=8$, $\alpha=0.65$), the formal potential ${E}^{0,\prime}=-0.257$ V on the same electrode-voltage grid, and the same measurement-error distribution (see Appendix \ref{sec:methods} for the full noise model). The four parameters $k$, $\tilde{\beta}_\mathrm{ox}$, $\tilde{\lambda}$, and $\alpha$ were fitted independently to each of the 50 noisy polarization curves. Thus, differences in the fitted parameter errors isolate the effect of the transport transition on the kinetic measurement.

\begin{figure}[htbp]
    \centering    
    \includegraphics[width=\linewidth]{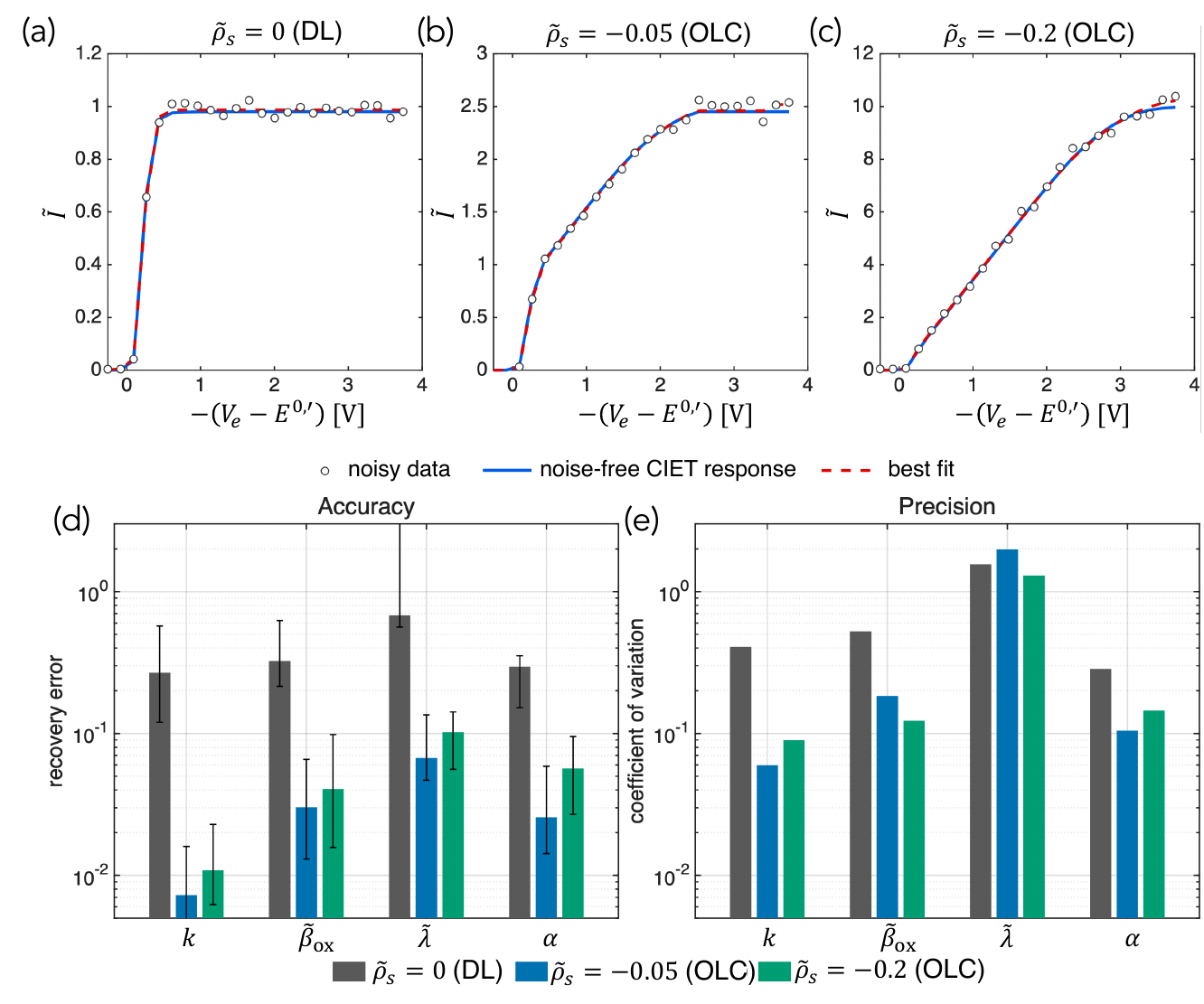}
    \caption{Charge transfer kinetic identifiability for the diffusion-limited case and two OLC-capable charged cases generated from a single set of reaction kinetic parameters ($Da=50$, $\tilde{\beta}_{\text{ox}}=8$, $\tilde{\lambda}=8$, $\alpha=0.65$, $\tilde{E}^{0,\prime}=-10$). Each of the 50 polarization curves contains $2\%$ relative Gaussian pointwise scatter together with curve-level gain, baseline, and drift contributions described in Appendix~\ref{sec:methods}. Polarization responses show one noisy replicate as circles, the exact CIET response in solid blue, and the best fit as a dashed red line for (a) $\tilde{\rho}_s=0$, (b) $\tilde{\rho}_s=-0.05$, and (c) $\tilde{\rho}_s=-0.2$. (d) Median relative recovery error of $k$, $\tilde{\beta}_{\text{ox}}$, $\tilde{\lambda}$, and $\alpha$ across the replicates with whiskers spanning the interquartile range. (e) Coefficient of variation of each fitted parameter across the same replicates, defined as $\operatorname{std}({\theta})/\operatorname{mean}({\theta})$. Lower values in panels (d) and (e) indicate greater recovery accuracy and precision, respectively.}
    \label{fig:CIET_identifiability}
\end{figure}

Fig.~\ref{fig:CIET_identifiability} (a--c) shows the simulated response under three surface charge conditions, where all cases share identical interfacial kinetics and differ only in surface charge. Each plot shows one noisy polarization curve (circles), the exact simulated CIET response (blue solid line), and the corresponding best fit (red dashed line). In Fig.~\ref{fig:CIET_identifiability} (a), the current saturates at the transport-controlled plateau for the uncharged case. In Fig.~\ref{fig:CIET_identifiability} (b) and (c), negative surface charge carries the response beyond the diffusion limit and toward the reaction-limited current. The fitted curves closely follow the calculated responses in all three cases, demonstrating that visual agreement alone is insufficient to establish whether the underlying charge transfer parameters are independently constrained. Thus, the distributions of fitted parameters should also be examined.

Fig.~\ref{fig:CIET_identifiability} (d) and (e) compare the accuracy and precision with which the CIET parameters are recovered in the uncharged diffusion-limited case and in two negatively charged cases that support OLC. Under diffusion limitation, the fitted CIET parameters vary widely even when the calculated and fitted polarization curves nearly coincide. Both negative-charge cases reduce the typical parameter errors by approximately one order of magnitude and strengthen the local sensitivity spectrum by more than one order of magnitude. The improvement is not monotonic with $|\tilde{\rho}_s|$, because identifiability is governed by the extent to which the sampled voltage window enters the portion of the polarization curve that remains sensitive to the charge transfer kinetics.

Among the fitted parameters, $\tilde{\lambda}$ remains the least well constrained and exhibits the largest coefficient of variation. In the ICET regime, $\tilde{\lambda}$ enters primarily through the effective kinetic prefactor $k_{\rm ICET}$, so changes in $\tilde{\lambda}$ can be compensated by changes in $k$. Once bulk salt is depleted, the interfacial cation concentration approaches $\tilde{c}_X|_L\approx-\tilde{\rho}_s$, and Eq.~\ref{eq:rxn_bc} gives $\tilde{I}_{\rm lim}\approx Da(-\tilde{\rho}_s)$. Thus, when $\tilde{\rho}_s$ and the transport properties are known independently, the overlimiting plateau directly constrains $Da$, and hence $k$, but provides no independent constraint on $\tilde{\lambda}$. Information on $\tilde{\lambda}$ therefore arises mainly from the curvature near the ICET--ECIT crossover. Surface conduction sustains counterion transport beyond bulk depletion and thereby makes this curvature more apparent in the polarization response; however, the limited voltage interval over which the crossover occurs and the electrolyte potential drop leave $\tilde{\lambda}$ strongly correlated with $k$.

\subsection{Application to Metal Electrodeposition}
\begin{figure}[htbp]
    \centering
    \includegraphics[width=\linewidth]{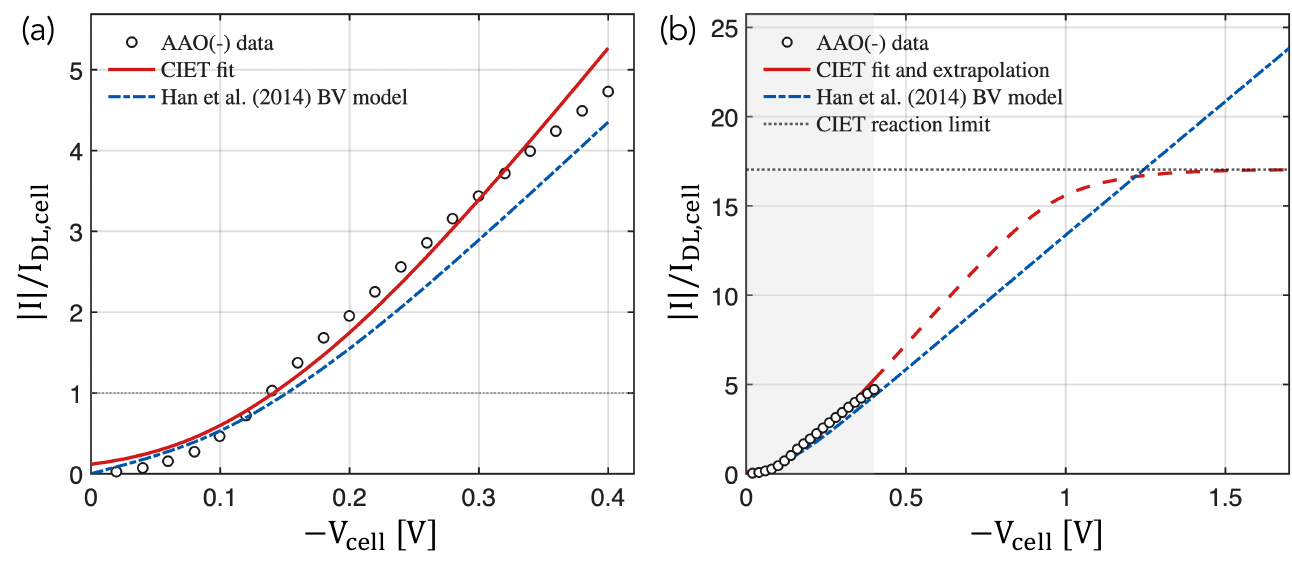}
    \caption{Cu electrodeposition through negatively and positively charged AAO membranes. (a) Polarization data from Ref.~\cite{han_over-limiting_2014} compared with the present CIET model and the BV model reported in that work. The CIET energy parameters are fitted to AAO($-$) and retained for AAO($+$), with only $\tilde{\rho}_s$ changed. Open circles and squares denote experimental data for AAO($-$) and AAO($+$), respectively. Red and blue curves represent the CIET and BV models, respectively, with solid lines corresponding to AAO($-$) and dashed lines to AAO($+$). (b) Model responses extrapolated beyond the measured voltage range. The shaded region indicates the measured voltage range shown in (a). The CIET model transitions from ICET-to-ECIT at $-V_{\rm cell}\approx0.12$~V for AAO($-$) and $0.11$~V for AAO($+$). At larger polarization, AAO($-$) approaches the finite reaction-limited current $\tilde I=Da(-\tilde{\rho}_s)\approx17$ (shown by the horizontal dotted line), whereas the BV response continues to increase.}
    \label{fig:han_fit}
\end{figure}

As discussed earlier, Han et al. \cite{han_over-limiting_2014} studied the problem of Cu electrodeposition in the overlimiting and underlimiting regimes in a system similar to the one we consider (Fig. \ref{fig:fig1}). The key difference is that instead of a reservoir, their cell had a Cu anode where Cu metal electrodissolves into Cu$^{2+}$ ions, so the current in the system is twice what would be obtained if a reservoir were in place of the anode---apart from that, the essential physics is the same. In their system, the I-V curve in the overlimiting regime appears to be linear at high voltage with no apparent plateau (Fig. \ref{fig:han_fit}), suggesting weak reaction limitations. Here, we will apply our theory to their data to quantitatively estimate the extent of reaction limitations posed by Cu electrodeposition in their system. 

We first note that for the case of the negatively charged anodized aluminum oxide (AAO) membrane, the maximum current obtained is 18.5 mA, which on scaling by the diffusion-limited current
$I_{\rm{DL}} = 3.9$ mA, yields $\tilde{I} \equiv \tilde{I}_{\rm{max}} = 4.7$. With no other side reactions, the limiting current must be greater than this value, so we must have $\tilde{I}_{\rm{lim}} >\tilde{I}_{\rm{max}} = 4.7$. Additionally, the authors estimate the surface charge due to the PSS coating on AAO as $-0.75$ e/nm$^2$, which after scaling yields $\tilde{\rho}_s = -0.71$. Then, using our analytical approximation in Eq. \ref{eq:Ilim_neg_rhos_large}, we can estimate a lower bound on the Damköhler number for Cu electrodeposition in their system.
\begin{equation}
    Da(-\tilde{\rho}_s) > 4.7 \implies Da > 6.7
\end{equation}
Since $Da>1$, the relation above already indicates that reaction limitations are weaker than transport limitations in this system. Indeed, we can go a step further and get a better estimate by additionally making use of their data on the limiting current obtained from the positively charged AAO membrane. The limiting current for AAO(+) is $\sim$ 2.3 mA, so that $\tilde{I}_{\rm{lim}} = 0.60$, and the estimated surface charge due to the PAH coating is 0.375 e/nm$^2$, yielding $\tilde{\rho}_s = 0.36$. Using our analytical approximation in Eq. \ref{eq:Ilim_pos_rhos}, we have
\begin{equation}
\frac{Da}{1+Da}\left[\exp\left(-\sinh^{-1}\left(\frac{\tilde{\rho}_s}{2}\right)\right) + \frac{\tilde{\rho}_s}{2}\left(\ln \tilde{\rho}_s - \sinh^{-1}\left(\frac{\tilde{\rho}_s}{2}\right)\right)\right] \approx 0.60 
\implies Da \approx 24
\end{equation}

The estimated value of $Da$ not only aligns with the lower bound derived, but confirms that reaction limitations are indeed much weaker than transport limitations in this system, supporting the authors' assumption of BV kinetics for Cu electrodeposition and electrodissolution. We can also estimate the reaction rate constant for Cu electrodeposition as
\begin{equation}
    k \equiv k_{\rm{CIET}}= Da \times \frac{ I_{\rm{DL}}}{zeAc_0} \sim 2.2 \times 10^{-4 }{\text{ m/s}}
\end{equation}

With $Da=24$ fixed from AAO($+$), the AAO($-$) curve was used to determine the two CIET parameters, $\tilde\beta_{\rm ox}$ and $\tilde{\lambda}$, in Eq.~\ref{eq:ciet_kinetics}. We assumed symmetric IT to set $\alpha=0.5$ for the assumed one-electron Cu$^{2+}$/Cu$^+$ rate-determining step~\cite{han_over-limiting_2014,bazant_unified_2023}. The fit to the AAO($-$) curve in Fig.~\ref{fig:han_fit} (a) yields $\tilde\beta_{\rm ox}=4.1$ and $\tilde\lambda=5.6$, with $R^2=0.98$. Without further adjustment, the same CIET parameters give $R^2=0.93$ for AAO($+$) after changing only $\tilde\rho_s$. Details of the data treatment are given in Appendix~\ref{sec:methods_han}.

Fig.~\ref{fig:han_fit} (a) compares the digitized AAO($-$) and AAO($+$) polarization curves from Ref.~\cite{han_over-limiting_2014} with the present CIET model and the BV model reported in that work. The fitted CIET model reaches the ICET-to-ECIT transition at \(-V_{\rm cell}\approx0.12\)~V for AAO($-$) and \(0.11\)~V for AAO($+$), so the higher-voltage polarization data lie on the ECIT branch. Nevertheless, the ECIT branch of CIET and the BV model give similar cell-level AAO($-$) polarization responses over the measured voltage range despite their different high-overpotential limits. For AAO($+$), the current is governed mainly by Cu$^{2+}$ exclusion and transport.

Fig.~\ref{fig:han_fit} (b) extends the model responses beyond the measured voltage. For AAO($-$), CIET approaches the finite reaction-limited current, $\tilde I=Da(-\tilde{\rho}_s)\approx17$, whereas the BV response continues to increase along the surface-conduction-supported branch. AAO($+$) remains underlimiting because positive fixed charge excludes Cu$^{2+}$ from the cathode region. Since ($Da\approx24$), the finite CIET plateau lies well beyond the measured current range that the present data do not directly test reaction-rate saturation. A more discriminating experiment would use a surface charge that is not only high enough to reach the OLC regime but also low enough to shift the reaction-limited plateau into an experimentally accessible voltage range.

Throughout both the measured and extrapolated ranges, the model assigns the Faradaic current exclusively to Cu electrodeposition and holds the membrane charge, active area, and deposit morphology fixed. Water splitting and deposit morphology could therefore alter the observed polarization response before the predicted CIET plateau is reached. Measurements designed to enter the reaction-limited regime, with these competing processes independently quantified or suppressed, would provide a more direct test of the finite reaction-limited current and stronger constraints on charge transfer kinetic parameters.

\section{Conclusion}
In this work, we developed a continuum model coupling the leaky membrane model to CIET kinetics at a reactive boundary to investigate the interplay of transport and reaction limitations. The coupled problem is governed primarily by two dimensionless parameters: the Damköhler number \(Da\), defined as the ratio of the reaction-limited current to the classical diffusion-limited current, and the scaled surface charge \(\tilde{\rho}_s\), which determines whether the reacting ion is enriched or excluded by the charged pore walls. Analytical approximations for the limiting current were derived in neutral, positively charged, and negatively charged porous media. For neutral media, the limiting current follows the series-resistance form \(I_{\rm lim}^{-1}=I_{\rm DL}^{-1}+I_{\rm RL}^{-1}\). Positive fixed charges suppress reactive-cation delivery and produce underlimiting currents, whereas negative fixed charges enrich reactive cations near the pore walls and can drive the system into the overlimiting regime. Importantly, overlimiting current is possible only when surface conduction is strong enough to overcome the reaction limitations, highlighting surface charge as a design parameter for tailoring the limiting current and polarization response.

In the overlimiting regime, when \(-\tilde{\rho}_s\gtrsim 2/(1+Da)\), the reactive-cation concentration near the electrode balances the fixed charge and the limiting current approaches \(I_{\rm lim}=I_{\rm RL}(-\tilde{\rho}_s)\). In this limit, the measured current directly reflects the intrinsic reaction capacity scaled by the amount of reactive charge supplied through surface conduction. This regime also preserves the sensitivity to charge transfer kinetics that would otherwise be hidden by diffusion limitation. Synthetic parameter-recovery studies show that overlimiting transport can substantially improve the accuracy and precision with which CIET parameters are inferred, although the reorganization energy remains difficult to identify unless the voltage window resolves the transition between IT- and ET-controlled regimes. Applying the model to published Cu electrodeposition data in polyelectrolyte-modified AAO membranes \cite{han_over-limiting_2014} yields \(Da\approx 24\), indicating that transport dominates the measured response. The fitted CIET parameters show that the behavior in the lower operating voltage range conforms to IT-controlled regimes (BV) while also clarifying that the same BV expression may not extend to the full experimental voltage range and should not be extrapolated indefinitely to larger overpotentials.

The present theory focuses on surface conduction as the mechanism enabling overlimiting current and, for simplicity, omits other effects that may be important in specific systems, including electroosmotic flow, electroosmotic instability, charge regulation, water splitting, evolving deposit morphology, and double-layer expansion at the reactive boundary. Incorporating these effects will be necessary for precise, quantitative device-specific models. Nevertheless, the reduced framework highlights two broader roles of charged porous media in electrochemical systems. First, the limiting-current map in the \((Da,\tilde{\rho}_s)\) plane provides design guidelines for tuning electrochemical response through surface charge, pore geometry, and electrolyte concentration. Second, by maintaining reactive-ion supply beyond the classical diffusion limit, charged porous media can reveal kinetic descriptors that are otherwise masked by concentration polarization. Thus, charged porous media act not only as transport layers but as tunable environments in which surface charge, ion transport, and charge transfer kinetics can be designed together to control limiting currents, investigate reaction kinetics, and shape macroscopic electrochemical performance.

\begin{acknowledgments}
A.V.Y. acknowledges support from the Center for Enhanced Nanofluidic Transport (CENT2), an Energy Frontier Research Center funded by the U.S. Department of Energy, Office of Basic Energy Sciences under Award \#DE-SC0019112. {J.Y. acknowledges support from the National Science Foundation under Award No. 2516269.} The authors thank Shakul Pathak for valuable insights on the Damk\"ohler number. 
\end{acknowledgments}

\appendix

\section{Numerical methods and identifiability analysis}\label{sec:methods}

\subsection{Solution of the coupled transport and reaction equations}\label{sec:methods_forward}

For each prescribed electrode potential $V_e$, the steady, one-dimensional transport equations of Section~\ref{sec2:mass_transfer} were coupled to the interfacial kinetic expression used in the calculation. Donnan equilibrium set the pore-side feed boundary, the nonreactive ions had zero flux, and the reactive-cation flux at $x=L$ was matched to the Faradaic reaction rate. Analytical integration then reduced the coupled problem to a nonlinear equation for the electrolyte potential at the electrode surface, $\tilde\phi_L$, obtained by enforcing electroneutrality at $x=L$. The equation was solved sequentially over $V_e$, and $\tilde I$ was evaluated from the integrated cation-flux relation.

For the monovalent synthetic calculations, the interfacial rate was evaluated at $\tilde\eta_f=\tilde{V}_e-\tilde\phi_L-\tilde E^{0,\prime}$. The same transport solution was used with each kinetic expression and only the interfacial rate expression was changed. The current scale $I_{\rm DL}$ and the reaction rate scale $Da$ are defined in Section~\ref{sec2:mass_transfer}. When a nonreactive supporting electrolyte was included, its ions were maintained in electrochemical equilibrium and included in the cross-sectional electroneutrality condition.

\subsection{Measurement-noise model and parameter estimation}\label{sec:methods_noise}

Each simulated measurement was constructed from the exact calculated current using prescribed CIET kinetics with pointwise scatter and curve-level instrumental errors added. The simulated current at voltage sample $i$ is
\begin{equation}
\tilde{I}_i^{\rm sim}=(1+g)\,\tilde{I}_i^{\rm CIET}+b+d\,x_i+\varepsilon_i,
\end{equation}
where $x_i\in[-1,1]$ denotes position within the potential sweep. The gain term $g$ had a standard deviation of $0.5\%$, while the sweep-wide baseline $b$ and linear-drift coefficient $d$ had standard deviations of $0.2\%$ and $0.3\%$ of the maximum current magnitude, respectively. The point-to-point term was normally distributed with
\begin{equation}
\sigma_i^2=(0.002\,\tilde{I}_{\max})^2+(0.02\,\tilde{I}_i^{\rm CIET})^2.
\end{equation}
Here $\tilde{I}_{\max}$ is the maximum magnitude of the exact current during the sweep. These values were chosen to represent relative current scatter, a small current-independent uncertainty floor, and sweep-wide gain, offset, and drift; they are not specifications of a particular instrument. Fifty perturbed polarization curves were analyzed for every condition in Section~\ref{sec:results_CIET}.

The CIET parameters were estimated by minimizing
\begin{equation}
    \chi^2 =\sum_{i=1}^{N_V}\left[\frac{\tilde I_i^{\rm model}-\tilde I_i^{\rm sim}}{\sigma_i}\right]^2,
    \label{eq:chi2_objective}
\end{equation}
where $\chi^2$ denotes the weighted sum of squared residuals and $N_V$ is the number of voltage samples. The weighting accounts for the current-dependent measurement scatter and prevents the high current portion of the polarization curve from dominating the fit. $\chi^2$ was only used as a fitting and relative model-comparison measure.

The positive parameters $k$, $\tilde{\beta}_{\rm ox}$, and $\tilde{\lambda}$ were fitted in logarithmic form, and $\alpha$ was constrained to $0<\alpha<1$. The formal potential $E^{0,\prime}$ was treated as known. Each four-parameter CIET fit was performed using six initial estimates, and the converged solution with the smallest $\chi^2$ was retained. 

\subsection{Sensitivity and uncertainty of the fitted kinetic parameters}\label{sec:methods_ident}

The sensitivity calculation asks whether changes in $k$, $\tilde\beta_{\rm ox}$, $\tilde\lambda$, and $\alpha$ produce distinguishable changes in the polarization curve under a specified transport condition. We refer to the ability to distinguish these individual kinetic contributions as parameter identifiability. At each fitted solution, the weighted sensitivity matrix was evaluated as
\begin{equation}
S_{ij}=\frac{\partial}{\partial\theta_j}\left[\frac{\tilde{I}_i^{\rm model}-\tilde{I}_i^{\rm sim}}{\sigma_i}\right]=\frac{1}{\sigma_i}\frac{\partial\tilde{I}_i^{\rm model}}{\partial\theta_j},
\end{equation}
where $\theta_j$ denotes $\ln k$, $\ln\tilde{\beta}_{\rm ox}$, $\ln\tilde{\lambda}$, and $\ln[\alpha/(1-\alpha)]$. For the singular-value calculation, the columns were normalized so that the result compared changes in polarization-curve shape rather than parameter units. A small minimum singular value indicates that two or more parameter changes produce nearly indistinguishable current responses, as occurs when diffusion limitation makes the cell current weakly dependent on interfacial kinetics.

Approximate local parameter intervals were obtained from
\begin{equation}
C_\theta=s_r^2(S^{\rm T}S)^{-1},
\qquad
s_r^2=\frac{\chi^2}{N_V-4},
\end{equation}
and mapped back to the physical parameter coordinates. These are local, linearized uncertainty estimates under the pointwise weighting in Eq.~\ref{eq:chi2_objective}. The simulated polarization curves also contain sweep-wide gain, baseline, and drift contributions, whereas the covariance calculation uses diagonal pointwise weights. The intervals therefore are not exact experimental confidence intervals. The variation among the 50 fitted parameter sets provides the complementary comparison reported in Fig.~\ref{fig:CIET_identifiability}.

\subsection{Analysis of copper electrodeposition data}\label{sec:methods_han}

The Cu/AAO/Cu polarization curves were digitized from Fig.~2(a) of Han \emph{et al.}~\cite{han_over-limiting_2014}. Then the extracted currents were normalized by the reported full-cell diffusion-limited current, $I_{\rm DL,cell}=3.90~{\rm mA}$, such that $\tilde I=|I|/I_{\rm DL,cell}$. The dimensionless surface charges, $\tilde\rho_s=-0.71$ for AAO($-$) and $+0.36$ for AAO($+$), were also used as reported in Ref.~\cite{han_over-limiting_2014}. The AAO($+$) current $2.32~{\rm mA}=0.595I_{\rm DL,cell}$ at high polarization was used as a proxy for the transport-limited response in Eq.~\ref{eq:Ilim_pos_rhos}, giving $Da\approx24$. This estimate depends on the prescribed charge and adopted transport approximation. It indicates that the reaction-current scale exceeds the diffusion-limited scale, but it does not identify the interfacial kinetic law.

For the CIET comparison, the two-electrode cell was represented by the reservoir-equivalent leaky-membrane calculation used in the theory. At each measured cell voltage ($V_\mathrm{cell}$), the model solved for the dimensionless electrolyte potential at the cathode, $\tilde\phi_L$, and evaluated the formal overpotential as
\begin{equation}
\tilde\eta_f=\tilde{V}_\mathrm{cell}-\frac{\tilde\phi_L}{z}, \quad z=2
\end{equation}
where $\tilde{V}_\mathrm{cell}=eV_\mathrm{cell}/k_BT$ is the cell voltage scaled by the thermal voltage. The completed electrodeposition current used $z=2$, whereas the assumed rate-determining Cu$^{2+}$/Cu$^+$ CIET step transfers one electron. The formal potential reference was set to zero for the symmetric Cu/AAO/Cu cell, and no voltage-offset parameter was fitted.

The symmetry factor was prescribed as $\alpha=0.5$ for symmetric IT~\cite{bazant_unified_2023}. The parameters $\tilde\beta_{\rm ox}$ and $\tilde\lambda$ were fitted only to AAO($-$), with bounds $0.05\leq\tilde\beta_{\rm ox}\leq200$ and $0.05\leq\tilde\lambda\leq100$. Residuals were weighted by $\sigma_i=[(0.05|\tilde I_i|)^2+(0.02\tilde I_{\max})^2]^{1/2}$. The fit gave $\tilde\beta_{\rm ox}=4.1$ and $\tilde\lambda=5.6$, with a weighted residual sum of $34.8$ and $R^2=0.98$. With the kinetic parameters unchanged, replacing only $\tilde\rho_s$ by the AAO($+$) value gave $R^2=0.93$ and a weighted residual sum of $107.5$. Uncertainty from digitization was examined by refitting 200 independently perturbed AAO($-$) curves, with $1.5~{\rm mV}$ voltage standard deviation and current standard deviation $\sigma_i$. All 200 fits were usable; the $2.5$th--$97.5$th percentile ranges were $3.08$--$4.96$ for $\tilde\beta_{\rm ox}$ and $4.23$--$7.14$ for $\tilde\lambda$. These values are conditional on the prescribed $Da$, $\alpha$, membrane charges, voltage mapping, current normalization, and transport properties.

The fitted calculation crosses the criterion $-\tilde\eta_f=\tilde\beta_{\rm ox}$ at $-V_{\rm cell}=0.123$~V for AAO($-$) and $0.109$~V for AAO($+$). The largest measured AAO($-$) current, $\tilde I=4.73$, is approximately $28\%$ of the extrapolated CIET reaction limit $Da(-\tilde\rho_s)=17.04$.

\section{Range of validity of BV and MHC approximations} \label{sec:results_approximations}
The calculations in this section use CIET as a prescribed reference response rather than assuming that CIET is the established kinetic mechanism of a particular electrochemical system. The purpose is to determine when the simpler BV and MHC expressions reproduce a CIET polarization curve within the assumed measurement uncertainty and when ionic transport preserves sufficient sensitivity to distinguish the underlying interfacial rate laws.

The ICET-dominant reference used $(\tilde{\beta}_{\rm ox},\tilde{\lambda},\alpha)=(30,5,0.5)$, whereas the ECIT-dominant reference used $(5,30,0.5)$. Both calculations used $Da=50$ and $\tilde E^{0,\prime}=-10$. The cathodic sweep depth is denoted by $-V_{e,\rm end}$, where $V_{e,\rm end}$ is the terminal electrode potential included in the fit. At each endpoint, 50 independently perturbed polarization curves were generated using the noise model in Section~\ref{sec:methods_noise}. The BV fits varied the apparent prefactor $k_{\rm BV}$ and cathodic transfer coefficient $\alpha_{\rm BV}$, whereas the MHC fits varied the apparent prefactor $k_{\rm MHC}$ and dimensionless reorganization energy $\tilde{\lambda}_{\rm MHC}$. The four-parameter CIET fit varied $k$, $\tilde{\beta}_{\rm ox}$, $\tilde{\lambda}$, and $\alpha$, with $E^{0,\prime}$ prescribed. Because $\eta_f=V_e-\phi_L-E^{0,\prime}$, the kinetic range sampled by a given applied-potential window also depends on the transport-dependent electrolyte potential $\phi_L$.

Fit quality was compared using a diagonal-weighted Akaike information criterion (AIC) score~\cite{akaike_new_1974,burnham_multimodel_2004}. For the fixed pointwise weights used here,
\begin{equation}
\mathrm{AIC}=\chi^2+2p,\qquad
\Delta\mathrm{AIC}_{\rm}
=\mathrm{AIC}_{\rm reduced}
-\mathrm{AIC}_{\rm CIET},
\label{eq:aic_diag}
\end{equation}
up to a model-independent constant, where $\chi^2$ is defined in Eq.~\ref{eq:chi2_objective} and $p$ is the number of fitted parameters: $p=2$ for BV or MHC and $p=4$ for CIET. Values near zero indicate that the polarization curves do not resolve the difference between the two expressions, whereas positive values indicate that the reduction in the CIET residual exceeds its additional parameter penalty.

\begin{figure}[htbp]
    \centering
    \includegraphics[width=\linewidth]{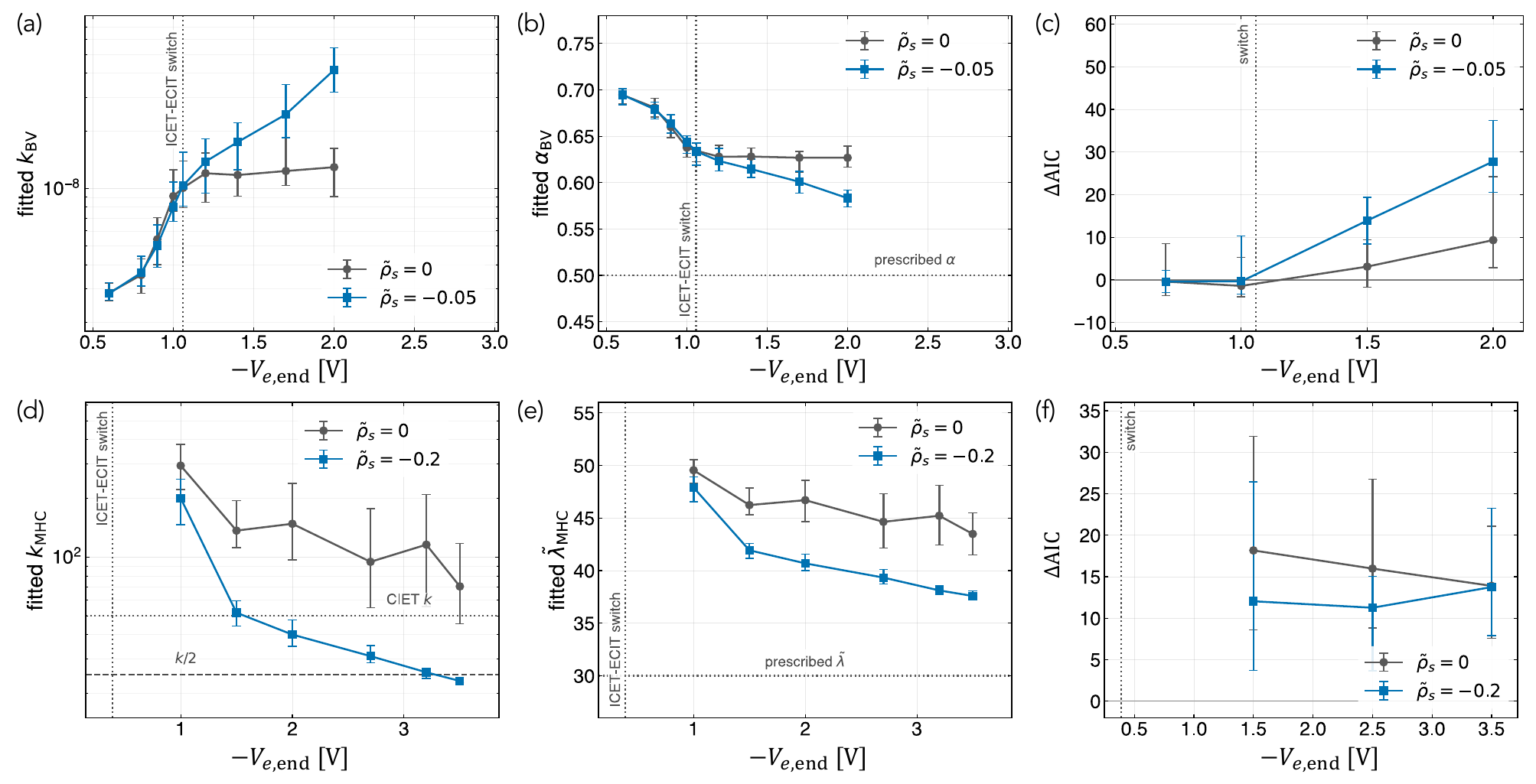}
    \caption{Dependence of reduced-model fitted parameter estimation and relative fit score on polarization-sweep depth at $Da=50$. Markers and whiskers denote the median and interquartile range over $N_{\rm rep}=50$ noisy curves. (a)--(c) BV fits to the CIET reference curves in the ICET-dominated regime for $\tilde\rho_s=0$ and $-0.05$: fitted (a) $k_{\rm BV}$, (b) $\alpha_{\rm BV}$, and (c) $\Delta{\rm AIC}$. (d)--(f) Corresponding MHC fits to the CIET reference curves in the ECIT-dominated regime for $\tilde\rho_s=0$ and $-0.2$: fitted (d) $k_{\rm MHC}$, (e) $\tilde\lambda_{\rm MHC}$, and (f) $\Delta{\rm AIC}$. Vertical dotted lines mark the ICET-to-ECIT transition. Horizontal dotted lines show the prescribed $\alpha$ in (b), CIET $k$ in (d), and prescribed $\tilde\lambda$ in (e); the dashed $k/2$ line in (d) denotes the implemented MHC normalization.}
    \label{fig:Fig5}
\end{figure}

Fig.~\ref{fig:Fig5} (a)--(c) shows that BV can reproduce the prescribed CIET response over a limited, approximately Tafel region preceding the ICET--ECIT transition. In that range, $\Delta\mathrm{AIC}$ remains near zero, so the observable current does not justify the additional CIET parameters. This response-level agreement does not make $k_{\rm BV}$ and $\alpha_{\rm BV}$ intrinsic CIET parameters. In particular, $k_{\rm BV}$ is an intercept extrapolated to zero formal overpotential and is strongly coupled to the fitted Tafel slope. For the OLC-capable case with $\tilde{\rho}_s=-0.05$, extending the sweep from $-V_{e,\rm end}=0.60$ to $2.0$~V increases the median $k_{\rm BV}$ by approximately 15-fold and decreases $\alpha_{\rm BV}$ from 0.695 to 0.583. The prescribed reaction and transport properties remain unchanged within this series; only the fitted potential interval changes. The parameter drift therefore reflects BV being forced to represent the curvature that develops beyond the ICET regime, rather than a physical change in the reaction. Correspondingly, $\Delta\mathrm{AIC}$ rises to approximately 14 and 28 at endpoints of $-1.5$ and $-2.0$~V in the charged medium, compared with approximately 9 at $-2.0$~V under diffusion limitation. BV is therefore a useful local approximation only while the interfacial response remains Tafel-like and its fitted parameters are stable to modest changes in the fitting interval. A diffusion-limited cell may conceal its failure outside that interval.

The MHC results in Fig.~\ref{fig:Fig5} (d)--(f) show the complementary requirement. MHC parameters become informative only when the measured response extends sufficiently far into the ECIT curvature and toward the reaction-limited current. For $\tilde{\rho}_s=-0.2$, increasing the sweep depth from $-1.0$ to $-3.5$~V moves the median $k_{\rm MHC}$ from 200 to 23.2 and $\tilde{\lambda}_{\rm MHC}$ from 47.9 to 37.6, while substantially narrowing their replicate distributions. In the implemented normalization, the MHC rate saturates at $2k_{\rm MHC}$, so the reference $k_{\rm MHC}=k/2=25$ is a prefactor convention rather than a microscopic prediction. The fitted $\tilde{\lambda}_{\rm MHC}$ nevertheless remains above the prescribed $\tilde{\lambda}=30$, because the two-parameter MHC law absorbs part of the finite IT contribution that CIET represents separately. Under diffusion limitation, the fitted parameters remain broad and nonmonotonic because increasing the applied voltage provides little additional information about the interfacial rate. The persistent $\Delta\mathrm{AIC}\approx11$--18 further shows that precise MHC parameters do not imply that the complete CIET response has been recovered.

\begin{figure}[htbp]
    \centering
    \includegraphics[width=0.8\linewidth]{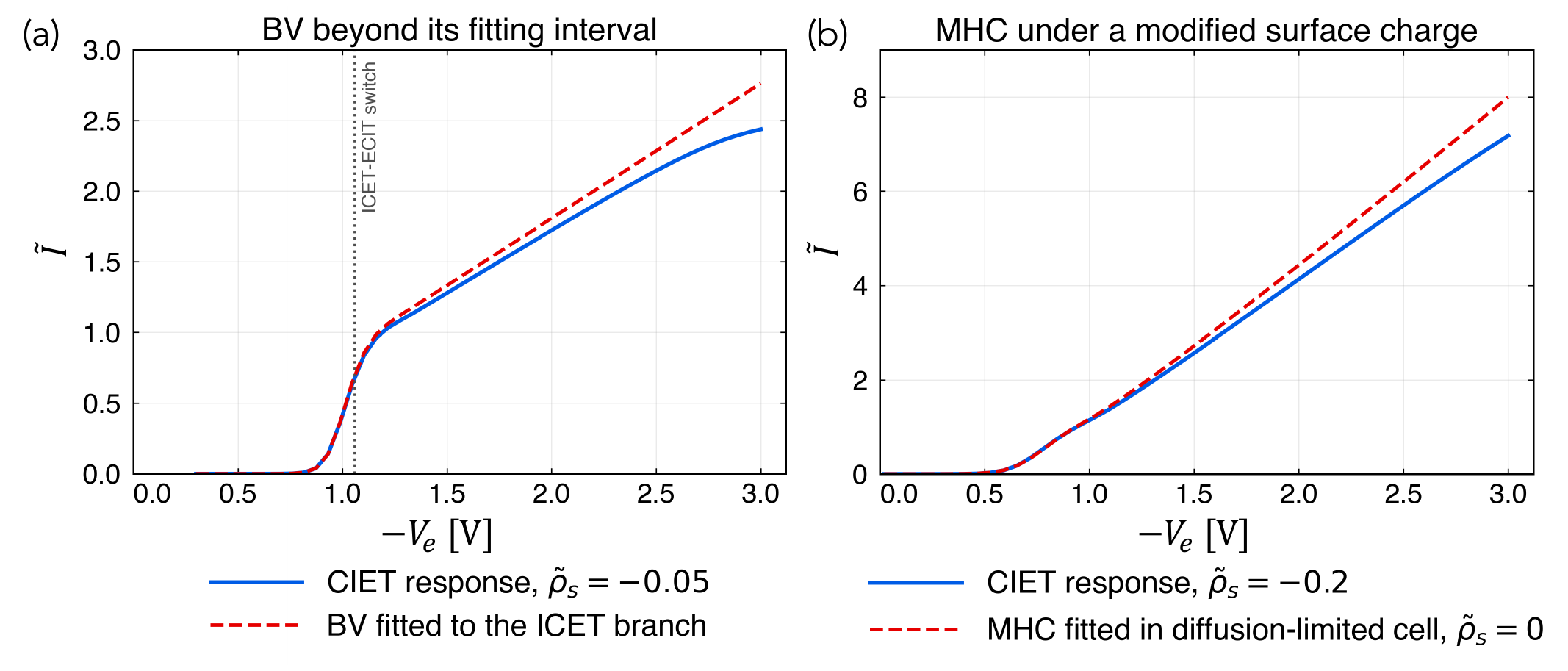}
    \caption{Transferability of reduced-model fitting over a specified potential range at $Da=50$. (a) A BV expression fitted on the ICET voltage window at $\tilde{\rho}_s=-0.05$ and extrapolated past the switch (vertical dotted line) reproduces the observable response to within $13\%$ out to $V_e=-3.0$ V, even though the same expression overestimates the interfacial rate at fixed formal overpotential by up to a factor of $4.9\times10^{3}$ over the same range. (b) An MHC expression calibrated in the diffusion-limited cell ($\tilde{\rho}_s=0$) systematically overpredicts the response of the OLC-capable device at $\tilde{\rho}_s=-0.2$, with the error growing from $7\%$ at $V_e=-2.0$ V to $11\%$ at $-3$ V.}
    \label{fig:Fig6}
\end{figure}

Fig.~\ref{fig:Fig6} demonstrates why agreement of the cell current is insufficient to validate an interfacial rate law. A BV expression fitted to the ICET voltage range at $\tilde{\rho}_s=-0.05$ overpredicts the cell current by only 13.4\% at $V_e=-3.0$~V, although at the same formal overpotential it overpredicts the prescribed CIET interfacial rate by a factor of $4.9\times10^3$. Coupled ion transport therefore masks most of the rate-law error in the observable current. In the MHC test, parameters fitted to the diffusion-limited response with $\tilde{\rho}_s=0$ through $V_e=-1.5$~V are then transferred to the OLC-capable response with $\tilde{\rho}_s=-0.2$ and extrapolated to larger polarization. The predicted current is 7.2\% high at $-2.0$~V and 11.4\% high at $-3.0$~V. This test therefore measures the combined cost of transferring parameters from a transport-masked condition and extrapolating beyond the calibration interval. Reduced parameters fitted under one transport condition should not be assumed transferable to another without such an out-of-condition test.

\begin{figure}[htbp]
    \centering
    \includegraphics[width=\linewidth]{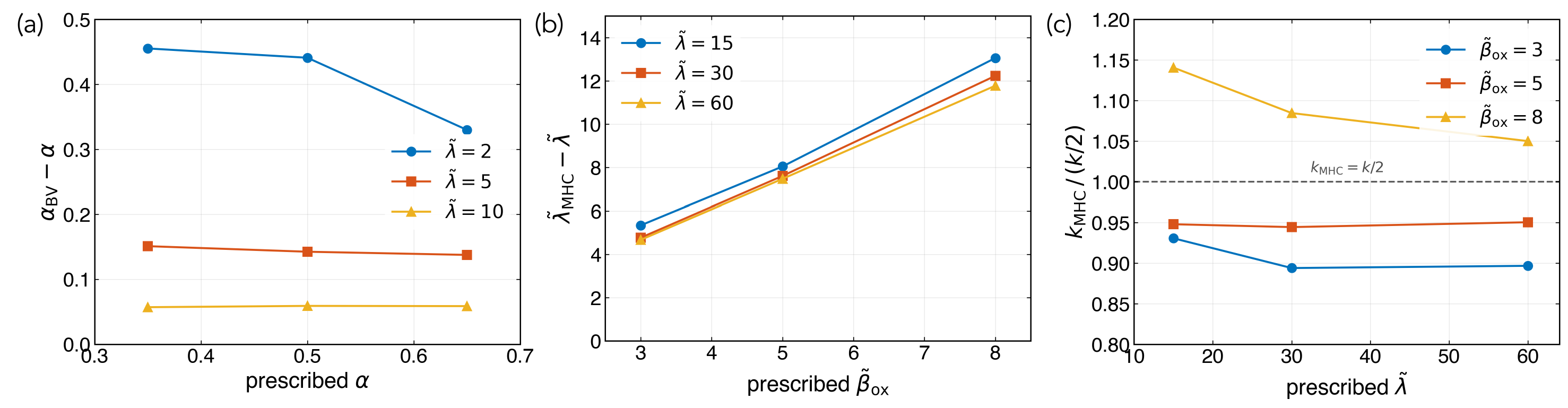}
    \caption{Biases from noise-free reduced fits at $Da=50$. The BV expression is fitted over the prescribed ICET voltage window at $\tilde\rho_s=-0.05$, and the MHC expression is fitted to the OLC response through $V_e=-3.5$ V at $\tilde\rho_s=-0.2$. (a) Shift in the fitted BV transfer coefficient; the $\tilde\lambda=2$ cases lack a distinct Tafel-linear branch. (b) Shift in the fitted MHC reorganization energy with the IT barrier. (c) Fitted MHC prefactor normalized by $k/2$, which records the implemented MHC saturation convention.}
    \label{fig:Fig7}
\end{figure}

The noise-free calculations in Fig.~\ref{fig:Fig7} confirm that these parameter shifts are systematic rather than consequences of measurement scatter. For BV, $\alpha_{\rm BV}-\alpha$ is approximately 0.14 at $\tilde{\lambda}=5$ and 0.06 at $\tilde{\lambda}=10$, while the $\tilde{\lambda}=2$ cases contain no distinct Tafel-linear interval within the prescribed fitting window. The fitted transfer coefficient therefore absorbs CIET curvature that is absent from the BV expression. For MHC, $\tilde{\lambda}_{\rm MHC}-\tilde{\lambda}$ increases from approximately 5 to 12--13 as $\tilde{\beta}_{\rm ox}$ increases from 3 to 8, showing that the apparent MHC reorganization energy also incorporates the omitted IT contribution. The ratio $k_{\rm MHC}/(k/2)$ remains between 0.89 and 1.14 over this grid. These relations depend on the selected potential range, transport condition, parameter grid, and weighting and should not be used as universal conversions between BV, MHC, and CIET parameters.

These calculations do not imply that BV or MHC is generally incorrect, nor do they establish CIET as the true charge-transfer mechanism. If an experimental reaction genuinely follows BV or MHC kinetics over the measured range, stable and transferable parameters should be recovered. The broader conclusion is that diffusion limitation can make distinct interfacial rate laws appear equivalent, whereas surface-conduction-supported OLC can preserve reactive-ion supply and expose differences in kinetic curvature and reaction-rate saturation~\cite{dydek_overlimiting_2011}. A discriminating experiment should therefore constrain $\tilde{\rho}_s$ and the transport properties independently, apply the same candidate kinetic parameters across several surface-charge conditions, test progressively wider potential intervals, and require prediction outside the calibration condition. Side reactions, water splitting, heating, charge regulation, and evolving electrode morphology must also be quantified before attributing high-field curvature or a plateau to charge-transfer kinetics. Under these controls, OLC measurements can reject inadequate kinetic expressions and more strongly constrain the interfacial mechanism, but agreement with CIET would constitute supporting evidence rather than proof of a unique mechanism.

\bibliographystyle{apsrev4-2}
\bibliography{refs_gpt_corrected}

\end{document}